\documentclass[twocolumn,aps,prl] {revtex4} 
\usepackage{graphicx}
\begin{document}
\pagestyle{empty} 
\title{Molecular dynamics study of contact mechanics: contact area and interfacial separation from small to full contact}
\author{C. Yang and B.N.J. Persson}
\affiliation{IFF, FZ-J\"ulich, 52425 J\"ulich, Germany}

\begin{abstract}
We report a molecular dynamics study of
the contact between a rigid solid with a randomly rough surface
and an elastic block with a flat surface. We study the contact area
and the interfacial separation from small contact (low load) to full
contact (high load). 
For small load the contact area varies linearly with the load and the interfacial separation
depends logarithmically on the load. For high load the contact area approaches to the nominal contact area
(i.e., complete contact), and the interfacial separation approaches to zero.
The present results may be
very important for soft solids, e.g., rubber, or for very smooth surfaces, where complete contact
can be reached at moderate high loads without plastic deformation of the solids. 

\end{abstract}
\maketitle


It is a very difficult to prepare surfaces which are really flat. Even on most
polished surfaces, hills and valleys are present which are large compared with the atomic-size.
Usually, if two solids are placed in contact, 
the upper surface will be supported on the summits of the irregularities,
and large surface areas will be separated by distances which are great compared with the molecular
range of action\cite{Bowden,BookP,Borri,Hyun1}. The separation
$u({\bf x})$ between the surfaces will vary in a nearly random way with the lateral
coordinates ${\bf x}=(x,y)$ in the apparent contact area. 
When the applied squeezing pressure increases, the contact area $A$ will increase and the average 
surface separation $u=\langle u({\bf x})\rangle$ will decrease, 
but in most situations it is not
possible to squeeze the solids into perfect contact corresponding to $u=0$. 
Understanding the area of real contact, and the interfacial separation between
two solids is essential to friction, adhesion, sealing and many other important applications\cite{Isral.book,confined.liq}.

Most studies of contact mechanics have been focused on small load where the contact area
depends linearly on the load\cite{Bush,GW, Hyun3, Chunyan,Carlos}. 
However, for soft solids, such as rubber or gelatin, or for smooth surfaces,
nearly full contact may occur at the interface, so it is of great interest to study how the
contact area, the interfacial surface separation and  stress distribution vary with
load from small load (where the contact area varies linearly with the load),
to high load [where the contact is (nearly) complete]. Here we will present such a study using
molecular dynamics(MD), and we will compare the numerical results with the prediction of the analytical
contact mechanics theory of Persson\cite{JCPpers,P1,preparation}.

\begin{figure}
\includegraphics[width=0.45\textwidth,angle=0]{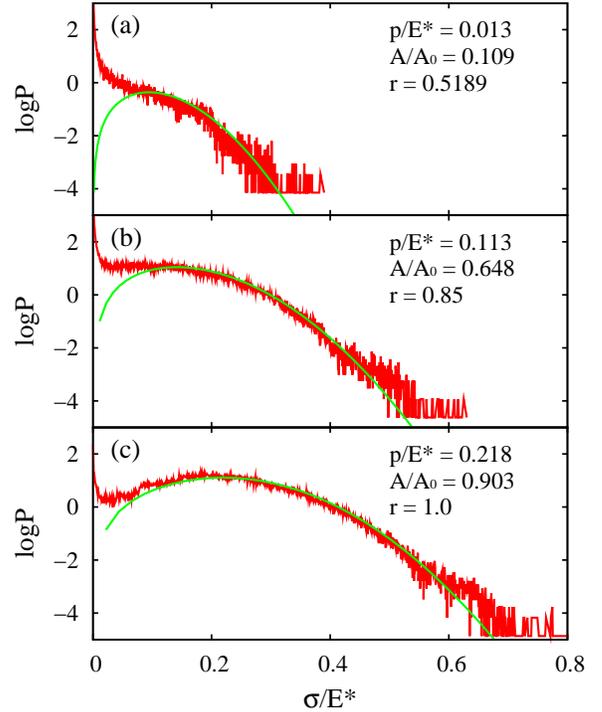}
\caption{\label{pressure_distribution}
The pressure distribution for $\zeta=4$ for three different nominal pressure, 
(a) $p/E^{*} \approx 0.013$,
(b) $p/E^{*} \approx 0.113$, (c) $p/E^{*} \approx 0.218$. 
The pressure probability distribution becomes broader with increasing the squeezing pressure. 
Properly choosing the correction factor (see Fig. \ref{block1}) makes the numerical results in
good agreement with Persson's contact mechanics theory.}
\end{figure}

Consider  randomly rough surfaces
with roughness wavelength components in some finite range 
$\lambda_1 < \lambda < \lambda_0$, where $\lambda_0$ is similar to 
(but smaller than) the lateral size of
the nominal contact area.
In order to accurately describe the contact mechanics between elastic 
blocks, it is necessary to consider solid block which extends (at least) a 
distance $\sim \lambda_0$ in the direction normal to the nominal contact area. 
This leads to an enormous number of atoms or dynamical variables even for a small 
systems. In order to avoid this trouble we have developed a 
multiscale MD approach\cite{Chunyan}.
The atoms at the interface between the block and substrate interact with the repulsive potential
$U(r)=\epsilon \left( r_{0}/r \right)^{12}$,
where $r$ is the distance between a pair of atoms, $r_{0}=3.28 \ {\rm \AA}$ and $\epsilon=74.4 \ {\rm meV}$.
In the MD-model calculations there is no unique way to define the separation $u$ between the
solid walls. 
Here we have used the same definition as in Ref. \cite{Chunyan} $u=d-d_{c}$, where $d$ 
is the difference between the plane through the center of the atoms of the top layer
of substrate atoms and bottom layer of block atoms.
$d_{c}$ is the critical atom-atom separation used to define contact on the atomic scale.
Thus, $u=0$ corresponds to the separation ${d_{c}=4.36 \rm \AA}$ between 
planes through the center of the interfacial atoms of the block and the substrate.

The system has lateral dimension $L_{x}=N_{x}a$
and $L_{y}=N_{y}a$, where $a$ is the lattice space of the block.
Periodic boundary condition is used in $xy$ plane.
For the block $N_{x}=N_{y}=400$, while the lattice space of the substrate
$b\approx a/\phi$, where $\phi=(1+\sqrt{5})/2$ is the golden mean, in order to avoid
the formation of commensurate structures at the interface. The mass of the block
atoms is 197 a.m.u. and $a=2.6 \ \rm \AA$, reproducing the atomic mass and
density of gold. The elastic modulus and Poisson ratio of the block is
$E=77.2 \ {\rm GPa}$ and $\nu=0.42$, respectively.
The substrate surface has self-affine fractal surface roughness\cite{P3,Chunyan}. 
For a self-affine fractal surface the power spectrum has power-law behavior
$C(q)\sim q^{-2(H+1)}$, where the 
Hurst exponent $H$ is related to fractal dimension $D_{\rm f}$ of the surface
via $H=3-D_{\rm f}$. For real surfaces this relation holds only for a finite wave vector region
$q_{0}<q<q_{1}$. 
Note that in many cases there is roll-off wavevector $q_{0}$ below which $C(q)$ is approximately constant.
The randomly rough substrates we use have been generated as described in Ref. \cite{P3}, and have
root-mean-square roughness $h_{\rm rms} = 10 \rm \AA$, fractal dimension is $D_{\rm f}=2.2$,
and roll-off wavevector $q_{0}=3q_{L}$, where $q_{L}=2\pi/L$. We define the
magnification $\zeta=q/q_{0}$.

\begin{figure}
\includegraphics[width=0.45\textwidth,angle=0]{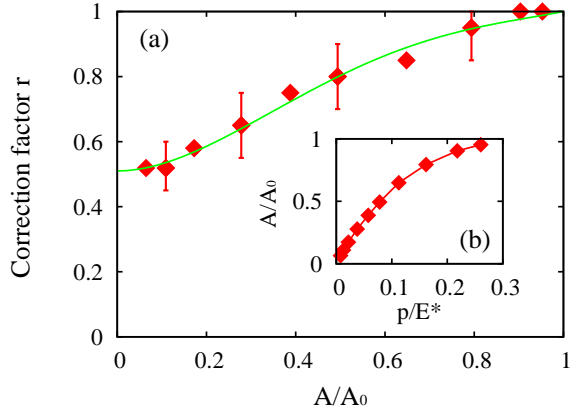}
\caption{\label{block1}
(a) Correction factor r as a function of the contact area ratio $A/A_{0}$. 
The points are simulation results, which have been fitted by the function
$f(x)=a+bx^{2}+cx^{3}+dx^{4}$ under the condition $d=1-a-b-c$ (solid line).
The corresponding coefficients $a$, $b$, $c$ are $0.51, \ 2.5, \ -3.3$ respectively.
(b) The contact area ratio $A/A_{0}$ as a function of squeezing pressure normalized by effective elastic modulus.}
\end{figure}

With molecular dynamics simulations we can calculate the interfacial stress distribution.
In order to obtain the contact area we follow the procedure outlined in Ref. \cite{Chunyan}
and fit the numerical results to the theoretically
predicted stress distribution
\begin{equation}
  \label{pres.dis}
 P(\sigma, \zeta) = {1\over 2 (\pi \tilde G)^{1/2}} 
 \left (e^{-(\sigma -p)^2/4\tilde G}-e^{-(\sigma +p)^2/4\tilde G}\right )
\end{equation}
where $\tilde G(p,\zeta)$ 
depends on the nominal squeezing pressure $p$ and the magnification $\zeta$
(but which is independent of $\sigma$, see below),
and which we choose to get the best fit with the numerical data.
In Fig. \ref{pressure_distribution} we have shown
the good agreement between the numerical pressure
distribution and the analytical theory 
(for $\zeta=4$) under three different nominal pressure.
Once $\tilde G$ is known we can calculate 
the relative contact area using\cite{PSSR}
\begin{equation}
  \label{area.inte}
  {A\over A_0} = \int_0^\infty  d\sigma \ P(\sigma,\zeta) 
\end{equation}
In Fig. \ref{block1}(b)  we show the relative contact area  $A/A_0$ 
as a function of normalized pressure $p/E^{*}$ from small to full contact.

The fitted $\tilde G(p,\zeta)$ can now be compared with theory. Thus, the theory
of Persson predicts $\tilde G = G$ where
\begin{equation}
  \label{G}
   G= {\pi \over 4} \left ( {E\over 1 - \nu^2}\right )^2 \int_{q_L}^{\zeta q_0}dq \ q^3 C(q) \,.
\end{equation}

Fig. \ref{block1}(a) shows the ratio $r=\tilde G/G$ which refers to correction factor.
Note that $r$ increases from $\approx 0.51$ to 1 as the squeezing pressure $p$
increases from  zero to infinite (i.e., the normalized contact area $A/A_0$ increases from zero to 1).
Since the contact area for small load is proportional to $\sim 1/\sqrt {G}$, it follows that the theory
for small load predicts a contact area about $\sim 30 \%$ smaller than that deduced from the MD simulation.
This is slightly larger than what has been found in earlier numerical simulations. Thus, the finite element
calculations of Hyun and Robbins\cite{Hyun2} and the Green's function molecular dynamics study of
Campana and M\"user\cite{Carlos} gives $r\approx 0.64$, corresponding to a contact area
about $\sim 20 \%$ larger than that predicted by the Persson theory. 
Similarly, the study of H\"onig\cite{Hoen} gives $r\approx 0.56$ for small load. 
However, none of the computer simulations
can be considered as perfectly converged, so the difference between theory and fully converged 
numerical simulation may be smaller than that indicated by the numbers given above. 

\begin{figure}
\includegraphics[width=0.45\textwidth,angle=0]{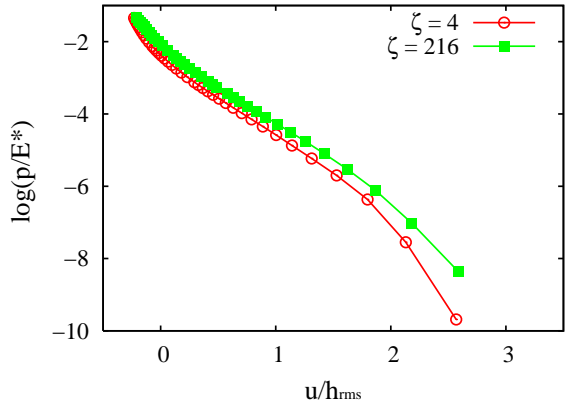}
\caption{\label{block}
An elastic block squeezed against a rigid rough substrate. The natural logarithmically normalized 
average pressure $log(p/E^{*})$, as a function of  normalized 
interfacial separation $u/h_{\rm rms}$ on different magnifications $\zeta=4$ and $\zeta=216$.}
\end{figure}

\begin{figure}
\includegraphics[width=0.45\textwidth,angle=0.0]{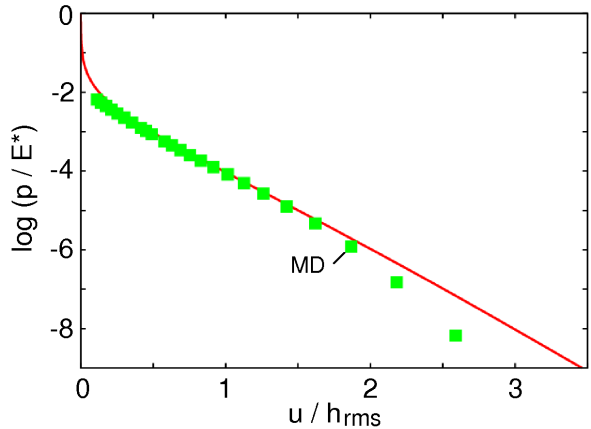}
\caption{\label{Chuny}
The relation between the natural logarithm of the
squeezing pressure $p$ (normalized by $E^{*}$) and the interfacial separation $u$ (normalized
by the root-mean-square roughness amplitude $h_{\rm rms}$)
for an elastic solid squeezed against a rigid surface. The theory curve (solid line) 
has been calculated using the theory presented in Ref. \cite{preparation} with
$\gamma = 0.42$.
}
\end{figure}

Recently Persson theoretically derived the relation between the average interfacial separation $u$
and the applied normal squeezing pressure $p$ \cite{preparation}. For nonadhesive interaction and small applied pressure,
$p \sim e^{-u/u_{0}}$, in a good agreement with recent experimental observations\cite{Benz}. 
Here we numerically calculate the average interfacial separation with different squeezing pressure with molecular dynamics.
In Fig. \ref{block} we show the natural logarithm of the normalized 
average pressure $p/E^{*}$, as a function of the normalized 
interfacial separation $u/h_{\rm rms}$. We show result for the magnification $\zeta=4$ (open circles)
and $\zeta=216$ (solid squares). Since the atoms interact with a long-range repulsive $\sim r^{-12}$ pair potential,
it is possible to squeeze the surfaces closer to each other than what corresponds to $u=0$.
This explains why simulation data points occur also for $u < 0$. 

In Fig. \ref{Chuny} we compare the MD results from Fig. \ref{block} (open circles) with the theory presented in
Ref. \cite{preparation} using the same surface roughness power spectra (and other
parameters) as in the MD-calculation. The theory is in good agreement with the numerical
data for $0.2 < u/h_{\rm rms} < 2$. For $u/h_{\rm rms} < 0.2$ the two curves differ because of the
reason discussed above, i.e., the ``soft'' potential used in the MD simulation allows
the block and substrate atoms to approach each other beyond $u({\bf x})=0$, while in the analytical
theory the potential is
infinite for $u ({\bf x})<0$ and zero for $u({\bf x}) >0$. The difference between the theory and the
MD results for  $u/h_{\rm rms} > 2$ is due to a finite size effect. That is, since the MD calculations use a 
very small system, the highest asperities are only $\sim 3 h_{\rm rms}$ 
above the average plane (see the height distribution in
Ref. \cite{details}), and for large $u$ very few contact spots will 
occur, and in particular for $u > 3 h_{\rm rms}$
no contact occurs and $p$ must vanish. In the analytical theory, the system size is assumed to be infinite.
Even for a Gaussian distribution of asperity height, there will always be infinitely many 
infinitely high asperities. Contact will occur at arbitrarily large separation $u$, and the
asymptotic relation $u\sim {\rm log} p$ will hold for arbitrarily large $u$ at small squeezing pressures
$p$.    

\begin{figure}
\includegraphics[width=0.45\textwidth,angle=0.0]{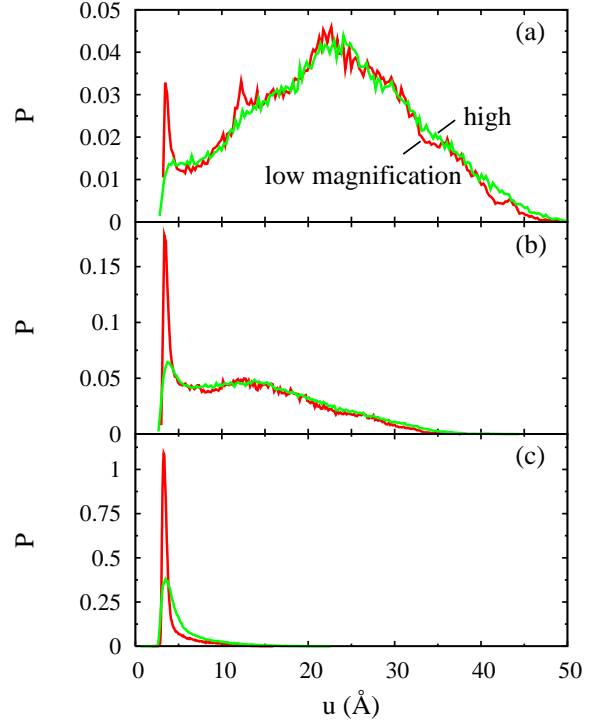}
\caption{\label{u.distribution}
Probability distribution of interfacial separation 
under different pressure (a) $p/E^{*} \approx 0.002$, 
(b) $p/E^{*} \approx 0.013$, (c) $p/E^{*} \approx 0.113$, at low and high 
magnifications respectively.
Note that the distributions of interfacial separations observed at low and 
high magnifications are similar for $u > 5 \rm \AA$. 
This result is expected since mainly the long-wavelength, large amplitude roughness will
determine the separation between the surfaces when the separation is large. }
\end{figure}

Let us now discuss the probability distribution of interfacial separation, defined by 
\begin{equation}
  \label{ave.def}
  P_{u}=\langle \delta(u-u({\bf x})) \rangle
\end{equation}
where $\langle...\rangle$ is ensemble average.
This function is shown in Fig. \ref{u.distribution} 
for three different loads and two different magnifications.
Note that the distributions of interfacial separations observed at low and 
high magnifications are similar for $u > 5 \rm \AA$. 
This result is expected since mainly the long-wavelength, large amplitude roughness will
determine the separation between the surfaces when the separation is large. 
The quantity $P_u$ has many important applications. For example, for lubricated contact
at low sliding velocity, one may estimate the contribution from shearing the liquid film
to the (nominal) frictional stress using
$$\sigma \approx \eta v \int_{u_{\rm c}}^\infty du \ {P_u\over u}$$
where $\eta$ is the viscosity and $v$ the sliding velocity, and where $u_{\rm c}$
is a cut-off separation of order nanometer (continuum fluid dynamics is not valid for
liquid films thinner than a few nanometers).
Another important application is for estimating the fluid leaking through 
sealing\cite{details}.

To summarize, we have performed a Molecular Dynamics (MD) study of the contact between an elastic
block with a flat surface and a rigid substrate with a randomly rough surface.
The interfacial pressure distribution agrees well with the analytical theory of Persson. 
We have also calculated the area of real contact and the interfacial separation
as a function of load 
from small to full contact,
and compared the
results with the analytical theory. For not too large or too small squeezing pressures,
the MD results show that
the interfacial separation $u$ depends logarithmically on the squeezing pressure 
$u\sim {\rm log} p$, in a good agreement with the analytical theory and with experimental observation.
Finally, we have studied the probability distribution of interfacial 
separation for different pressure and different magnifications.
The present results may be of great importance for soft solids, 
e.g. rubber-like material, or very smooth surfaces.

\vskip 0.3cm

{\bf Acknowledgments}
We thank Ugo Tartaglino for many useful discussions related to the MD-simulation.

\end{document}